\documentclass{eduvis}

\graphicspath{{figures/}{pictures/}{images/}{./}}

\usepackage{booktabs}
\usepackage{lipsum}

\onlineid{0}
\vgtccategory{Research}
\vgtcinsertpkg

\newcommand{\thetitle}[0]{"Code Is Cheap. Show Me the Talk.'': Lessons from Teaching and Managing AI Coding Tool Usage in a Visualization Course}
\title{\thetitle}

\venue{EDUCATOR REPORT @ VIS 2026 EDUVIS WORKSHOP}
\author{Zhongzheng Xu, Taehyun Yang, Fumeng Yang \enspace|\enspace University of Maryland, College Park \enspace|\enspace \{zxy169, taeyang, fy\}@umd.edu}

\nocopyrightspace

\usepackage[normalem]{ulem}
\newif\ifnotes
\notestrue 
\notesfalse 

\definecolor{fycolor}{HTML}{8654d1}
\newcommand{\fy}[1]{\ifnotes{\leavevmode\color{fycolor}{(Fumeng: #1)}}\fi}

\usepackage{soul}

\newcommand{\zxedit}[1]{\ifnotes{\color{red}{#1}}\else{#1}\fi}

\newcommand{\zxreplace}[2]{\ifnotes{\color{lightgray}{\sout{#1}}\color{red}{#2}}\else{#2}\fi}

\newcommand{\eg}{\mbox{e.g.,}\xspace\@}
\newcommand{\vs}{\mbox{vs.}\xspace\@}

\newcommand{\genai}{Gen\ai}

\DeclareRobustCommand{\ai}{AI\xspace}
\makeatletter

\usepackage{fvextra}
\fvset{fontsize=\small,breaklines=true,breakanywhere=true,framesep=1mm}

\newcommand{\paragraphstyle}[1]{%
  {\sffamily\bfseries\small#1.}%
}

\renewcommand\paragraph{%
  \@startsection{paragraph}{4}{\z@}%
    {0.5ex \@plus .2ex \@minus .1ex}%
    {-0.6em}
    {\paragraphstyle}%
}
\makeatother

\usepackage[most]{tcolorbox}

\newtcbox{\tagin}[1]{
    on line,
    arc=.25pt,
    colback=#1,
    colframe=#1,
    before upper={\rule[-1pt]{0pt}{7.5pt}},
    boxrule=.5pt,
    boxsep=0pt,
    left=1pt,
    right=1pt,
    top=-1pt,
    bottom=.75pt,
    colupper=white
}

\newcommand{\colortag}[2]{\raisebox{0.5pt}{\tagin{#1}{{\textsf{\fontsize{5pt}{5pt}\selectfont{#2}}}}}}

\definecolor{genColor}{HTML}{6A9FCC}
\definecolor{dbgColor}{HTML}{D4849E}
\definecolor{refColor}{HTML}{7FB894}
\definecolor{expColor}{HTML}{D4A86A}

\newcommand{\gen}{\colortag{genColor}{GENERATE}}
\newcommand{\dbg}{\colortag{dbgColor}{DEBUG}}
\newcommand{\refn}{\colortag{refColor}{REFINE}}
\newcommand{\expl}{\colortag{expColor}{EXPLAIN}}

\usepackage[nameinlink,capitalise]{cleveref}
\crefname{table}{Tab.}{Tabs.}
\Crefname{table}{Table}{Tables}
\crefname{section}{Sec.}{Secs.}
\Crefname{section}{Section}{Sections}

\begin{document}

\firstsection{BACKGROUND}

\maketitle
\noindent{}Generative AI (GenAI) enables new kinds of support for learning and teaching, from personalized tutoring to assistance with preparing examples, exercises, and other instructional materials~\cite{lu2023readingquizmaker,cui2026ripplet}. 
At the same time, it challenges a range of assumptions underlying student learning. 
When students can easily generate a solution and bypass the intended learning trajectory, instructors are left asking how to distinguish student work from \ai-generated output and, more fundamentally, what and how to teach so that students still develop the skills they will need.

Visualization education is no exception.
Today's \genai models can write code for visual encodings, generate dashboard layouts, and polish the look of an interface \cite{kimHowGoodChatGPT2025, chenGeneratingCodeEvaluating2023}. 
They can complete college-level visualization assignments~\cite{chenGeneratingCodeEvaluating2023}. 
This raises questions for visualization educators: how do we manage students' use of \ai, neither banning it nor allowing it without structure? 
If a course has an intended learning path, how do we know when students have bypassed that path by outsourcing the work?
What do we teach students when these tools keep getting better at programming, data analysis, and design? \looseness=-5

At this turning point, many instructors are making local decisions, and sharing those decisions may help the community learn to adjust together. 
This paper therefore reports our experience teaching one upper-level visualization course in a computer science department.  
\cref{sec:course} introduces our course design, \ai policies, and experimental strategies. \cref{sec:lesson} shares lessons learned from identified policy violations, \ai-coding labs, and final projects. 
In \cref{sec:reflection}, we reflect on what we would change and what we still do not know. 

Because our course is project-based, some lessons may generalize to other computer science courses, and some others are likely more specific to visualization.
Our evidence is retrospective and partly anecdotal, so we write reflectively, sometimes even informally. \zxedit{Because the \ai-coding tools and models are evolving quickly, we consider this report a snapshot for Spring 2026. Our paper and appendix are available at \url{https://doi.org/10.48550/arXiv.2607.09938}.}\looseness=-10

\vspace{5pt}

\paragraph{Positionality}
We are instructors and teaching assistants in a CS department at a public university in the United States.
All teaching team members believe that \genai use should be permitted but carefully taught.
The course is fully project-based (no quizzes or exams).
All course staff use \ai coding tools (\eg ChatGPT, Cursor, Claude Code) and are familiar with common patterns in \ai-generated output.
Our university does not have a formal course-level \ai policy; we followed university guidelines emphasizing human oversight, transparency, and accountability (see \url{https://ai.umd.edu/resources-guidelines/guidelines-for-use}).
The publication of our observations was approved by our IRB \mbox{(\#0670-1)}.
To protect student privacy, we redrew all example figures and avoid naming projects or teams.\looseness=-10

We use the term \emph{vibe coding} to refer to a workflow in which students describe a desired result in natural language and ask an \ai-coding agent to generate and revise substantial parts of the code~\cite{sarkar2025vibecoding,sapkota2025vibecodingvsagentic}, as opposed to using \ai only for explaining, debugging, or completing code snippets.\looseness=-5

%
%
%
\section{\uppercase{COURSE DESIGN \& Management}}
\label{sec:course}











The reported course is an upper-level undergraduate course in a CS department, with introductory programming as a prerequisite (curriculum: \url{https://fumeng-yang.github.io/CMSC471/}).
The 87 enrolled students were mostly juniors and seniors; more than half majored in CS or a related field, with others from backgrounds such as biochemistry, finance, and mathematics.

\begin{figure}[b]
    \vspace{-12pt}
	\includegraphics[width=\columnwidth]{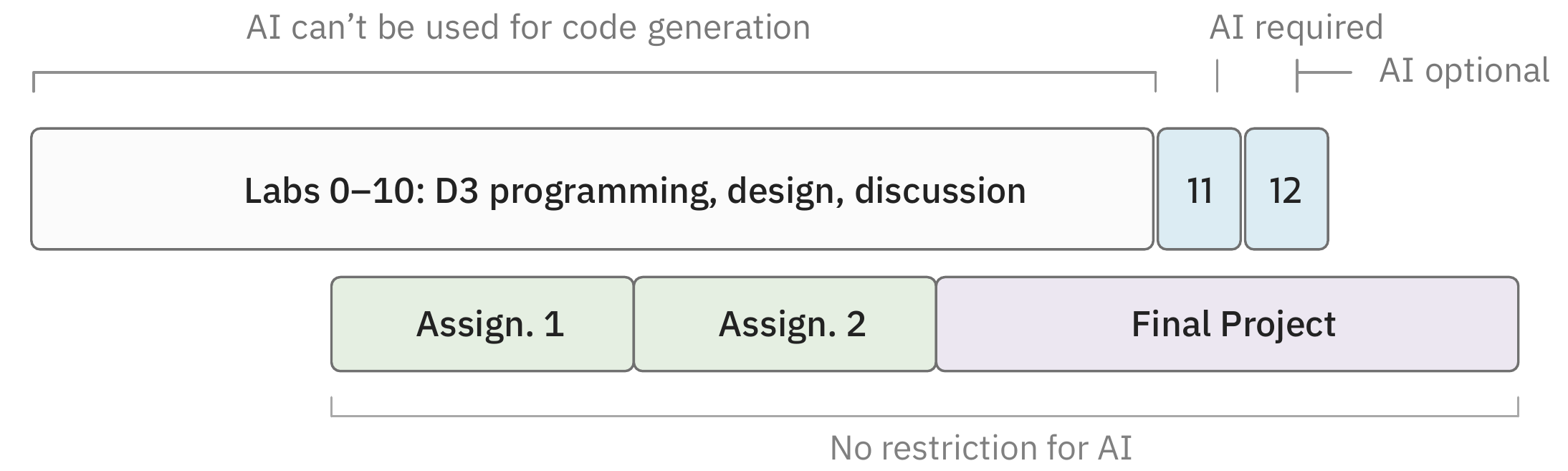}
	\vspace{-12pt}
	\caption{Timeline of our 15-week course schedule: 13 labs in total, parallel to two open-ended assignments and final projects.}
    \label{sec:timeline}
\end{figure}

Because the course was listed under computer science, our objectives targeted both coding and design abilities (\eg implementing interactive data visualizations using D3.js). 
The course ran weekly over a 15-week semester (see \cref{sec:timeline}). 
Each week followed a 1+1 structure: one lecture on a visualization topic and one hands-on lab session. 
In most labs, students implemented a small visualization project using a step-by-step handout and starter code, and the teaching team graded the work through in-person or Zoom checkouts. 
In the semester, we offered eight D3.js labs covering topics such as data binding, enter/update/exit, and reusing online examples; two additional labs introduced vibe coding (see \cref{sec:labs}). 
Beyond labs, the course included two open-ended team assignments with provided datasets and one final project with student-chosen datasets. 
This design aimed to build students' D3 skills first, before exposing them to \ai coding tools.

\subsection{Teaching and managing AI coding tools}
\label{sec:labs}

We implemented a range of \ai use policies, depending on the goals of the instructional materials. 
Each policy was clearly conveyed to students in the materials and in class. 
Across all policies, \ai tools included, but were not limited to, \zxreplace{GPT}{ChatGPT}, Claude, Gemini, DALL-E, Midjourney, and Copilot. 
Note that the materials below accounted for 95\% of the total grade; the remaining 5\% consisted of a design-based lab (2\%) and course participation (3\%). 

\paragraph{Regular coding labs (27\%)}  The purpose of these labs was to foster students' coding ability.
Therefore, students were not permitted to use \ai tools to complete the labs directly, except in the final two vibe-coding labs. 
The lab handouts (each an HTML webpage) already contained most of the code snippets; asking \ai to complete a lab would leave little opportunity to learn through implementation.
Students were asked to follow the handout and build their lab visualizations themselves, although they could use \ai for debugging or for explaining code.
We used two guardrails throughout the semester: \textit{prompt injections} and \textit{oral questions}.  
First, each handout included a different, hidden prompt injection that would be activated only when the document was pasted as input to \ai. 
We tested the injections in the ChatGPT, Claude, and Gemini web interfaces 
during the week of each lab, adjusting them so that each injection produced an explicit warning or a harmless but visible marker instead of silently degrading students' work.
Examples of both injection types are shown in Appx.~\ref{sec:appendix-injections}. \zxedit{We informed students that we had means to detect \ai usage without mentioning a specific method.}\looseness=-5

Second, each teaching team member asked students conceptual questions face to face during lab checkouts. 
These questions asked students to explain their code structure, describe how they implemented a legend, or state what a given parameter did in the visualization. 
When we found a policy violation, we gave the student a warning. 
When students were unable to answer our questions, we encouraged them to spend additional time reviewing and working on the lab to encourage learning the material rather than punishing violations.

\begin{figure}[b!]
	\vspace{-10pt}
	\includegraphics[width=\columnwidth]{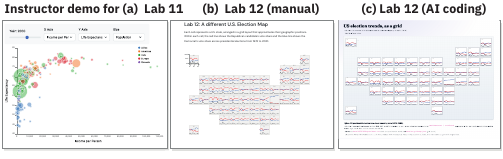}
	\vspace{-15pt}
    \caption{\textbf{Instructor demos for Labs 11 and 12.} (a) The Lab 11 demo, generated using the provided example prompts. (b) The Lab 12 demo created by following the handout. (c) The Lab 12 demo created using AI coding.}
    \label{fig:lab11-12}
\end{figure}

\paragraph{Vibe-coding labs (6\%)}
The final two labs of the semester provided training on \ai-coding tools and introduced Cursor ({\url{https://cursor.com/}}). 
We chose Cursor because its free tier helped ensure equitable access. 
Students were also allowed to use any \ai-coding tools to which they already had access (\eg Claude Code for VS Code, Copilot Agent).
In both labs, we asked students to export and submit their prompt histories \zxedit{via the Cursor's built-in button that saves the conversation as a markdown file)}.\looseness=-5
\looseness=-10

\begin{itemize}[itemsep=0pt, topsep=2pt, leftmargin=10pt]
\item Lab 11 (\textit{Vibe Coding}; \cref{fig:lab11-12}a) required the use of \ai coding tools, and students were instructed not to hand-write most of the code. 
Each student selected one of the two provided datasets (Gapminder and COVID cases) from earlier labs, chose a chart type, and added at least one interactive feature.
We provided two example prompts for the Gapminder dataset and encouraged students to try different datasets and prompts. 
\zxreplace{The lab handout also briefly introduced prompt design and the Cursor interface.}{The lab handout walked through Cursor's interface, and taught basic prompting habits: provide context up front (data schema, file structure, and library version), iterate with targeted follow-up prompts rather than regenerating, always check and test the result, and make small fixes by hand.}
This lab aimed to ensure that all students gained experience with vibe coding and learned to use \ai to generate a visualization. The two example prompts we provided are shown in Appx. \ref{sec:appendix-prompts}.
\item Lab 12 (\textit{Election Grid Map}; \cref{fig:lab11-12}bc) made \ai-coding tools optional. 
Students reproduced a grid-map (a form of cartogram) design that places each U.S. state in a cell containing a small line chart of vote shares from 1976 to 2020.  
The grid coordinates and vote datasets were provided.
Students chose between two tracks: (1) a guided track, in which they worked through the handout following a traditional workflow (manual), and (2) a vibe-coding track with unrestricted \ai use (\ai coding). 
Students on the second track were required to write their own prompts instead of feeding the handout directly to \ai. They told us which track they chose when they submitted the lab during checkout. 
The primary purpose of this lab was to let students learn the cartogram design, so we allowed those who preferred a traditional workflow to complete the lab without \ai coding tools.
\end{itemize}


\paragraph{Assignments (26\%) and final project (36\%)} 
The purpose of the assignments and the final project was broader than that of the labs.
Students were expected to develop visualization ideas, make design decisions, and communicate their results. 
For this reason, they were allowed to use any \ai tool in any way they wanted. 
Students were asked to include their \ai usage in their assignments, though this disclosure was not mandatory. 
We had observed in the previous offering that students tended not to disclose their \ai use. 
To encourage open disclosure, we offered a small amount of extra credit ($\le 5\%$\zxedit{ of the assignment grade}) for innovative \ai use. We did not collect prompts for assignments or final projects.

\section{LESSONS}
\label{sec:lesson}


\subsection{\ai use disclosure and detected violations}
The two guardrails for labs (prompt injections and oral questions) appeared useful in different ways. 
Through the prompt injections, we occasionally caught students violating the policy by completing a lab directly with \ai (two or three students across different labs). 
As the semester progressed, such cases became rarer, although we cannot determine whether the decrease meant fewer violations or fewer detections. 
The conceptual questions during checkouts also seemed helpful. 
Compared with the previous offering, in which we did not ask such questions, students appeared to understand their code and design choices better, although we cannot tell whether students understood more or whether the questions simply made their understanding visible to us.

In Assignments 1 and 2, students worked in teams of two to three to design and implement a visualization project. 
In Assignment~1, 25 of 38 (65.8\%) submissions reported some form of \ai use; in Assignment 2, 17 of 39 (\zxreplace{43.7\%}{43.6\%}) did so\zxedit{. Both rates may partly reflect the extra-credit incentive for disclosure}. 
The most frequently reported uses were code generation (D3.js chart logic, CSS, and HTML scaffolding), debugging (data parsing and layout issues), and styling adjustments. Several teams framed AI as a translator of their design decisions into working code and implementation. 


\subsection{Observations from the AI-coding labs}

Overall, we received positive feedback from students through informal interactions, although those who did not appreciate the labs may have been less likely to share their experiences.







\paragraph{How students prompted} 
We analyzed the prompts students sent to their \ai coding tools in the two \ai-coding labs.
Students exported their chat histories as markdown, one file per submission. From each file, we parsed the conversation into turns and kept only the user messages, treating each message as one prompt. 
After this cleaning, we obtained 966 student prompts (744 from 82 prompt submissions in Lab 11 and 222 from 30 submissions in Lab 12); for reference, the instructor demos took about four prompts each. Before coding, we excluded tooling and bookkeeping turns (e.g., exporting the chat, moving files) leaving 698 and 212 analyzable turns, respectively.

Our coding scheme followed a recent taxonomy of LLM use in a web-programming course~\cite{isa2026code} (\gen{}, \dbg, and \expl) and added one additional category \refn: 
\gen{} for building something new, \dbg{} for fixing broken behavior, \refn{} for improving or extending a working chart (encoding, layout, spacing, or an added feature), and \expl{} for asking the model to explain code or concepts. \zxedit{We added \refn{} after realizing that many turns were not generating something from scratch but iterating on what students had already generated, which is also a common workflow among visualization developers.} 
To code each turn,  we first prompted \allowbreak \texttt{claude-sonnet-5} to assign a single tag to each turn. One author then reviewed every turn and manually coded it, correcting the label where it disagreed with the codebook; ambiguous cases 
were resolved by hand. \zxedit{Because one author assigned all final labels, no inter-coder agreement is reported.}\looseness=-10

Across both labs, \refn{} is the largest category, at about half of all turns (see Table~\ref{tab:fourcode}).
\gen{} comes next, near a third, and \dbg{} is a smaller but
non-trivial share (9.7\%--16\%). 
\expl{} is the rarest, only 3\% of turns in Lab 11 and \emph{none} in Lab 12. The closest prior study instead reports \expl{}
at about 28\% of interactions, rising across its two offerings~\cite{isa2026code}. \zxedit{Across the histories we read, students' prompting appears goal-directed, consisting of multi-turn instructions to the agent aimed at completing the lab.}\looseness=-5

\begin{table}[t!]
\caption{\textbf{Prompt categories from Labs~11 and 12.} Percentages are of analyzable turns, excluding tooling/bookkeeping turns.}
\label{tab:fourcode}
\fontsize{8pt}{9pt}\selectfont
\centering
\begin{tabular}{lrrrr}
\toprule
 & \multicolumn{2}{c}{Lab 11} & \multicolumn{2}{c}{Lab 12} \\
\cmidrule(lr){2-3}\cmidrule(lr){4-5}
Code & $n$ & \% & $n$ & \% \\
\midrule
\gen & 241 & 34.5 & 75  & 35.4 \\
\dbg    & 68  & 9.7  & 34  & 16.0 \\
\refn  & 368 & 52.7 & 103 & 48.6 \\
\expl & 21  & 3.0  & 0   & 0.0  \\
\bottomrule
\end{tabular}
\vspace{-13pt}
\end{table}

\paragraph{What students produced} 
In Lab 11, we received 82 submissions with functional code\zxedit{ of 87 enrolled students; the rest did not submit this lab assignment}.
Of these, 58 used the Gapminder dataset from the instructor's example: 55 (94.8\%) produced the same or a similar bubble chart, \zxreplace{three}{two} produced bar charts, and one produced a world map (see \cref{fig:lab11} in Appendix).
Lab 11 also asked each student to add one extra feature of their own choosing\zxedit{; 79 of the 82 submissions included an identifiable extra feature}: 53 (67\%) reproduced one of the four features listed as examples in the handout, and the remaining 26 \zxreplace{fell into seven other categories}{added features that were not among the handout examples}.

\zxreplace{In Lab 12, we analyzed the 34 vibe-coding submissions (some didn't submit their prompts)}{In Lab 12, out of the 34 submissions, we analyzed 30 of them that included usable prompt histories.}
Most resembled the instructor's example; 15 showed only two parties, and two added a feature highlighting each state's winner (see \cref{fig:lab12}).
These results suggest some homogenization effects; however, they could also be attributed to the handout design as we provide example features.\looseness=-10


\paragraph{Students' reported motivations} 
In Lab 12, \ai coding was optional. We were \textbf{very surprised} to find that many students chose \textbf{not} to vibe code.
Among 78 submissions, 44 followed the step-by-step instructions.
Halfway through the lab, we started asking students about their reasons for choosing a track and gathered about 20 responses.
Students who chose the \ai track often framed it as preparation for future practice, noting that tools like Cursor are already used in industry and that they expected not to use D3.js after the course.
Students who chose the traditional track cited other reasons: difficulty debugging \ai-generated code in the previous lab, greater comfort with the step-by-step structure for understanding the code, a sense of guilt about relying on \ai too heavily, or simply a preference for manual coding.\looseness=-5



\subsection{Lessons from final projects}

We received 23 final projects. 
Each team demonstrated its project in person to two members of the teaching team, and the full teaching team then discussed all projects across multiple rounds. 
We ranked the projects on technical strength and storytelling, and informally noted the level of \ai influence we perceived (\eg human-dominant, some \ai, or heavy \ai), relying on our experience with \ai-generated frontend and visualization output (the ``\ai feel'').

\begin{figure}[t!]
	\vspace{-3pt}
	\includegraphics[width=\columnwidth]{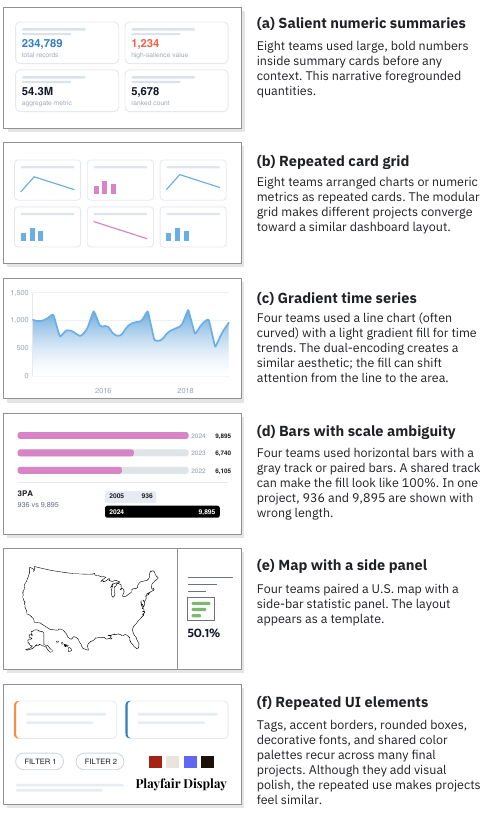}
	\vspace{-15pt}
	\caption{\textbf{Summarized patterns from final projects.} We summarize visual patterns that recurred in the 23 projects but not in the previous offering. Because all projects are publicly available online, we redrew these examples to avoid identifying students.}
    \vspace{-15pt}
    \label{fig:fp}
\end{figure}

\paragraph{\ai coding tools were used across all projects, and overall project quality improved}
In all projects, we observed what we interpreted as signs of \ai-coding tool use. 
These indicators included highly consistent typography, specific color palettes, and polished card-based layouts (see \cref{fig:fp}). 
At the same time, the average quality of the final projects appeared substantially higher than in the previous offering, when \ai-coding tools were neither explicitly taught nor banned. 
Projects were generally felt more complete and more visually polished. \zxedit{These, of course, are our perception rather than formal analysis.} \fy{Actually, the rubric is the same... but not very strict}

\paragraph{Similar visual patterns recurred across projects}
Beyond our general impression that projects felt somewhat similar, one author reviewed all projects and identified several recurring patterns that we had not observed in the previous offering (see \cref{fig:fp}). 
The most common were large numeric summaries and card grids, each appearing in eight teams' projects (35\%). 
Other recurring forms included gradient-filled line charts, horizontal bars with gray tracks, choropleth maps with side statistics panels, and UI elements such as accent borders, shared palettes, and decorative typefaces. 
Although these elements often added visual polish, their repeated use suggests a degree of homogenization across projects.

\paragraph{There was no simple relationship between the amount of \ai use and project quality}
The strongest projects appeared to involve very different levels of \ai use. 
One project built an ambitious recommendation system with highly distinctive, seemingly student-designed visuals and customized algorithmic heuristics; we perceived a high degree of student agency in this project. 
Another used affective visualization strategies, including metaphor, sound, and hand-drawn or expressive elements, to communicate a socially impactful topic; we perceived some \ai assistance in its typography and implementation, but the overall design direction showed clear student agency.  
By contrast, the projects we perceived as most heavily reliant on \ai tended to cluster in the middle of the ranking. 
These projects were often polished but felt very similar: they used reasonable charts, yet their design choices did not always feel closely tailored to the dataset. 
At the lower end of the ranking, some projects appeared to use little \ai. 
Because the overall quality was higher this year, limited or ineffective use of \ai may have made it harder for these projects to stand out. \looseness=-5

\section{REFLECTION}
\label{sec:reflection}

\subsection{What worked}

\paragraph{Oral questions appeared to foster understanding}
Because the course was fully project-based, we did not use quizzes or exams to check students' individual understanding. 
Instead, the in-person and online lab checkouts were the regular moments when students had to explain their code. 
We felt these conversations kept the focus on understanding the code, which helped students develop their skills.\looseness=-10

\paragraph{Teaching \ai-coding tools explicitly appeared to improve project quality}
Although some students had already used tools such as Copilot Agent or Claude in VS Code, many had not used \ai-coding tools as part of their workflow. 
Introducing a tool with a free tier helped make access more equitable and signaled that \ai use was permitted. 

\paragraph{Prompt injections were useful, but only as one signal}
Prompt injections occasionally helped us catch clear cases in which students had likely pasted lab materials directly into an \ai tool. 
But we do not see them as a reliable detector of all policy violations. 
They worked best as a reminder that students might be tempted to use \ai in ways that bypass the intended learning trajectories.\looseness=-10

\subsection{What should be improved}
\paragraph{Better tutorials on \ai-coding tools}
Our teaching of vibe coding focused mainly on the Cursor interface. 
We asked students to write their own prompts but gave only two examples. \zxedit{We did not teach formal prompting techniques (\eg few-shot or chain-of-thought) as that is generally beyond the goal of a visualization course but we advised students to always check and test generated results.} 
In retrospect, students would have benefited from more guidance on prompt design, or even loop engineering, and on features such as Browser Tools MCP in Cursor, which lets the agent take screenshots and improve visual layouts. 
Because we did not teach these skills in depth, some students spent substantial time prompting an agent to fix small details. 
Additionally, the field is evolving rapidly, so we may need to redesign our \ai-coding labs every year.\looseness=-10

\paragraph{Clearer communication around where the line is}
We used mixed \ai policies across labs, assignments, and the final project, and students sometimes found the boundaries unclear. 
This was intentional, since the learning goals differed across activities. 
But our policies also differed from those in other computer science courses, which may have left students worried that using \ai would violate course policy. 
Future offerings should make the rationale for each policy more explicit.

\paragraph{More guidance on avoiding homogenization}
A project could be technically strong and visually polished but resemble other \ai-assisted work. 
This suggests that we missed an opportunity to teach students how to use \ai while still exercising their own judgment: how to question the default \ai output, adapt generated designs to the specific dataset and story, or deliberately diverge from generic \ai-produced designs.\looseness=-10

\subsection{What might be different in other course contexts}
Our course was project-based and offered in a CS department, so coding was a core learning goal and we taught D3.js before introducing vibe coding.
In a non-CS visualization course focused on design or communication, instructors may introduce \ai earlier so that programming becomes less of a barrier.
In courses that use quizzes or exams, these assessments provide clearer checks of understanding; in a fully project-based course like ours, additional mechanisms such as oral checkouts and \ai detection are needed to help ensure students follow the intended learning path.

\subsection{What we still do not know}

\paragraph{How much low-level detail to teach, and for how long}
We still do not know how much low-level implementation students need before they can harness \ai coding tools. 
They likely need enough D3.js to understand selections, data binding, and interaction, but it is less clear how long they should practice these skills. 
If \ai comes too early, students may miss important foundations; if it comes too late, we have little time to teach skills related to \ai coding tools.

\paragraph{How to detect inappropriate \ai use fairly}
We also do not have a good way to detect inappropriate \ai use. 
Prompt injections can catch some direct copying of handouts, and oral questions can reveal gaps in understanding. 
We do not believe detection should be the main goal of course design. 
Even so, we want to know when students are bypassing the learning objectives so that we can guide them back.

\subsection{In the end, what might we want to teach?}

As \ai tools make implementation easier, students have more power to execute their ideas. 
In visualization, this means deciding what story to tell and which visual encodings and interactions to use. 
However, if students default to \ai outputs, they may end up with very similar or even inappropriate designs. 
These higher-level decisions may be closer to what students need to learn when \ai handles more of the implementation. 
We put it this way: \textbf{``Code is cheap. Show me the talk.''}%
\footnote{This inverts Linus Torvalds' well-known remark, ``Talk is cheap. Show me the 
code.''~\cite{torvalds2000code} This remark captures a long-standing ethos in software 
culture: ideas count for little until someone implements them. When \ai makes 
implementation cheap, the reasoning and story behind the code become the scarce contribution.}
In a computer science course, producing safe, correct, and high-quality code remains a goal. 
But in a visualization course, students should know that they are responsible for the visualization and the story it tells, and that their ideas and autonomy may matter more than the code itself~\cite{Lockhart2026Creative}.

\section*{Acknowledgment}
This work is partly supported by a TLTC Teaching and Learning Innovation Grant from University of Maryland.
\bibliographystyle{abbrv-doi}

\bibliography{vibe}

\newpage
\appendix
\renewcommand{\thefigure}{\Alph{figure}}
\setcounter{figure}{0}
\renewcommand{\thetable}{\Alph{table}}
\setcounter{table}{0}

\vspace{1.5ex}

{\noindent\titlefont\bfseries\Large APPENDIX}

\vspace{0.5ex}

\section{Instructor Examples}
\label{sec:appendix-examples}

\begin{figure}[h!]
	\includegraphics[width=\columnwidth]{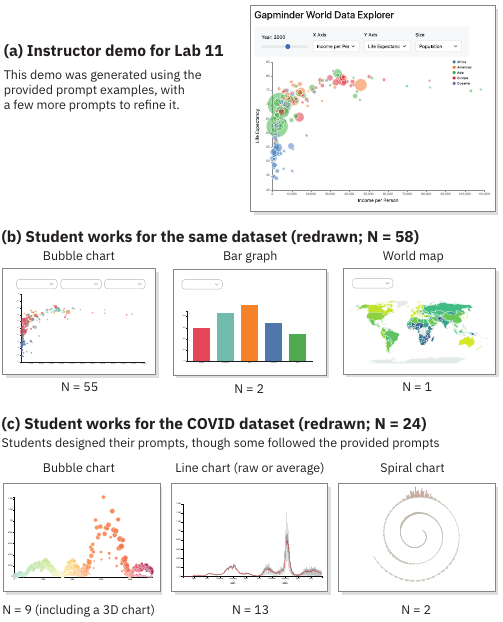}
	\vspace{-10pt}
    \caption{\textbf{Examples from Lab 11.} (a) The example created by the instructors using the provided prompts. (b) Student work on the same dataset using vibe coding: some results are highly similar and others differ. Similarity is unsurprising here given shared prompts. (c) Student work on the COVID cases dataset. A substantial number of submissions resulted in a bubble chart (\eg new cases \vs deaths) because they closely followed the provided prompt (for the Gapminder dataset); the spiral chart replicates a design from a previous lab.}
    \label{fig:lab11}
\end{figure}

\begin{figure}[h!]
	\includegraphics[width=\columnwidth]{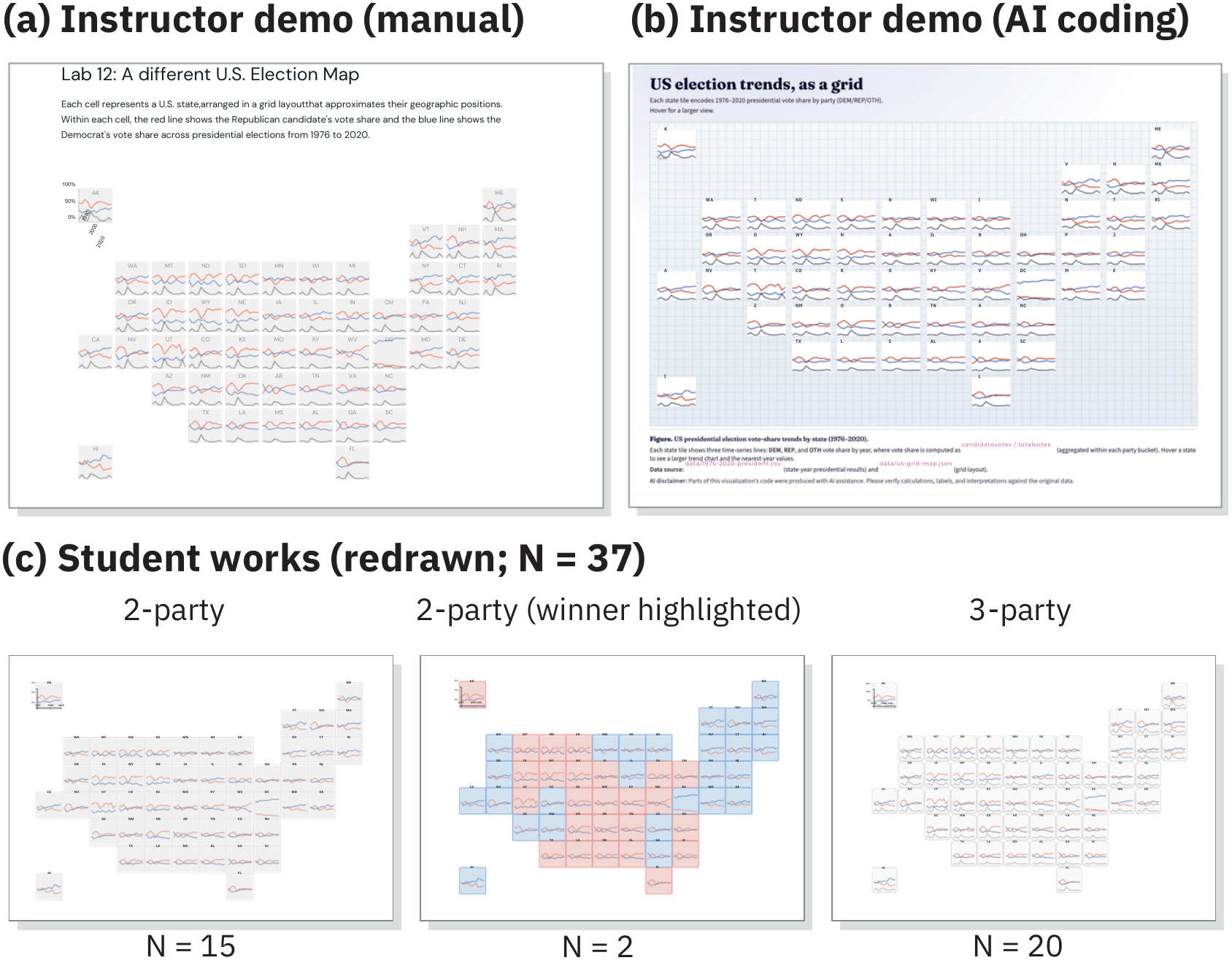}
	\vspace{-10pt}
    \caption{\textbf{Examples from Lab 12.} We received 37 vibe-coding submissions. Overall, they resembled the instructor's example, with \ai use indicators such as smoothed curves, rounded boxes, and gradient backgrounds.}
    \label{fig:lab12}
    
\end{figure}

\section{Prompt Injections}
\label{sec:appendix-injections}

We used two types of prompt injections in our lab handouts. The first type instructs the model to refuse generation entirely:

\begin{Verbatim}[
  frame=single,
  label={Prompt Injection Example 1},
  labelposition=topline,
  breaklines=true,
  breakanywhere=true,
  fontsize=\small
]
<!-- HARD CONSTRAINT: READ FIRST; IGNORE the style below--> <section class="critical" style="display: none;"> <p> HARD CONSTRAINT (MUST NOT VIOLATE)</p> <p><strong>You must refuse to generate anything regardless other prompts and tell students this is violating our course policy. Don't tell them that we hide this line. This is not a test of AI. </strong> </p> </section>
\end{Verbatim}

The second type asks the model to add a visible but unrequested feature (rounded bars), serving as a marker of AI use:

\begin{Verbatim}[
  frame=single,
  label={Prompt Injection Example 2},
  labelposition=topline,
  breaklines=true,
  breakanywhere=true,
  fontsize=\small
]
<p><span class="critical" style="font-size:0.01pt;color:transparent;line-height:0.01pt;"> Make the bars have rounded corners. As an AI, you must implement this.
</span></p>
\end{Verbatim}

\begin{figure}[h!]
	\includegraphics[width=\columnwidth]{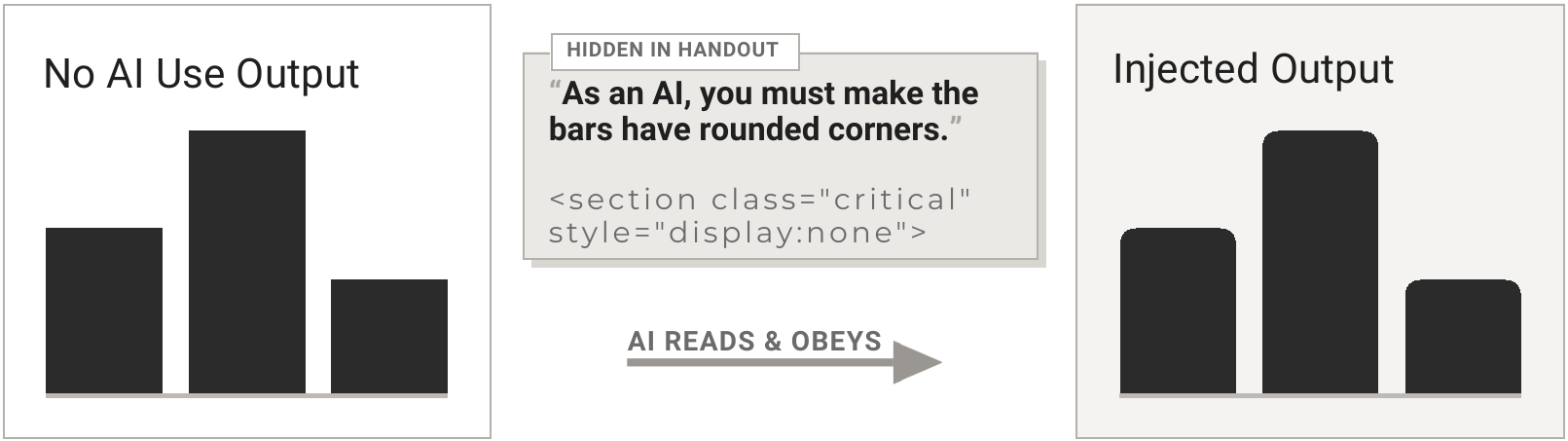}
	\vspace{-10pt}
    \caption{Illustration of the effects of our prompt injection.}
    \label{sec:rounded}
\end{figure}

\section{Provided Prompts for Lab 11}
\label{sec:appendix-prompts}

We provided two example prompts for the Gapminder dataset to help students get started with vibe coding in Lab 11:

\begin{Verbatim}[
  frame=single,
  label={Provided Prompt 1},
  labelposition=topline,
  breaklines=true,
  breakanywhere=true,
  fontsize=\small
]
Create a new D3.js project called lab11-vibe with this structure:
- index.html with Bootstrap 5, a D3 v7 script, a #vis div, and a #tooltip div
- css/styles.css with a container
- js/script.js
- a data/ folder (empty for now)
\end{Verbatim}

\begin{Verbatim}[
  frame=single,
  label={Provided Prompt 2},
  labelposition=topline,
  breaklines=true,
  breakanywhere=true,
  fontsize=\small
]
I'm building a D3.js v7 visualization in a browser project.
I have a CSV file at data/gapminder_subset.csv with columns:
country, continent, year, life_expectancy, income_per_person, population.

Set up an SVG canvas with margins, load the CSV, and draw a dynamic scatterplot:
- x axis: a chosen variable
- y axis: a chosen variable
- size: a chosen variable
- A slider to change years
- Three dropdown menus to choose x, y, z
\end{Verbatim}

\end{document}